# Electrically Detected Magnetic Resonance Applied to the Study of Near Surface Electron Donors in Silicon

W.D. Hutchison[a], P.G. Spizzirri[b], F. Hoehne[c] and M.S. Brandt[c]

[a] *Centre for Quantum Computer Technology, School of PEMS, The University of New South Wales, ADFA, Canberra, ACT 2600, Australia.*

[b] *Centre for Quantum Computer Technology, School of Physics, The University of Melbourne, Parkville, Victoria 3010, Australia.*

[c] *Walter Schottky Institute, Technical University of Munich, D-85748 Garching, Germany.*

Electrically detected magnetic resonance (EDMR) is applied to mm size devices with implanted leads and a 50 μm square gap laid down on bulk phosphorus doped silicon. Devices with a range of phosphorus concentrations and surface types were prepared and measured to examine the interplay between donor and charge trap states in producing EDMR signals.

## 1. Introduction

Magnetic resonance of donors in semiconductors via electron spin resonance (ESR) is well established. In particular the ESR of shallow electron donors in silicon was first achieved in the 1950's [1]. However, the sensitivity of conventional ESR is limited, requiring samples with $10^{10}$ donors or more. This problem can be overcome by detecting magnetic resonance via the effects of spin selection rules on other observables, such as charge transport.

Electrically detected magnetic resonance (EDMR), is where a change of the dc conductivity due to spin resonance is observed. EDMR was first demonstrated on Si:P by Schmidt and Solomon [2]. More recently, McCamey *et. al.* [3] showed that EDMR could be used to detect as few as 50 spins in a submicron size silicon device into which the phosphorus donors had been implanted. EDMR is also particular useful in the study of surface defects on semiconductors and their influence on donors placed near to the surface. In the case of shallow donors (eg phosphorus) in silicon (Si:P), it has been proposed that the spin dependent recombination/scattering of the photoelectrons proceeds via a process also involving (deeper energy) surface electron traps like the so called $P_b$ silicon interface dangling bonds [4].

In this paper we describe the development of a robust multi-micron EDMR device in silicon, with a view to detailed comparisons of the effects of different surface preparations, as well as variations in donor profiles. Preliminary results using bulk doped substrates with native and thermal oxides, as well as H- and D-terminated surfaces are presented and discussed.

## 2. Experimental details

A series of bulk doped, Si:$^{31}$P EDMR devices with various surface terminations were prepared using optical photolithography techniques. Three doping densities, $3 \times 10^{15}$, $2 \times 10^{16}$ and $1 \times 10^{17}$ P cm$^{-3}$ were used together with the following four different surface types (i) native oxide, (ii) high quality thermal oxide, (iii) H- and (iv) D-terminations (SiH & SiD). A microscope image of a typical device is shown in Fig. 1. The gap between the leads is 50 μm square and the buried metallic (highly doped) leads were created via low energy ion implantation of P$^+$. An RTA anneal was applied post implantation of the leads to repair damage. Large evaporated Al ohmic contacts allow direct coupling of current leads without the need for wire bonding to external contacts. The device design minimises the number of process steps and allows the metal contacts and external leads to be kept away from the microwave active region in the ESR cavity. ALl surfaces were prepared prior to implantation and annealing by cleaning (piranha and RCA2) with a subsequent removal of the pre-existing native oxide using hydrofluoric acid (HF). Controlled 5 nm thick thermal oxides were grown at 820°C. The H (D) terminated surfaces were prepared by etching the native oxide after processing with a 5% HF in $H_2O$ ($D_2O$) reagent. These later surfaces were preserved by a final covering with photoresist.

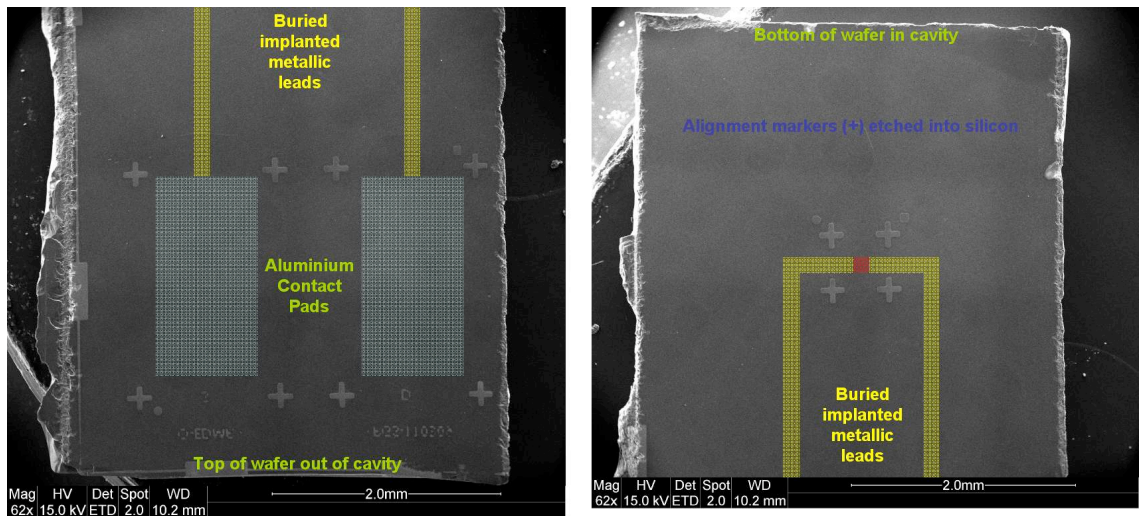

Fig. 1. Scanning electron microscope images of the upper and lower halves of an EDMR device.

The EDMR apparatus at the WSI utilises a standard x-band ESR system to provide the microwaves, dc and modulated magnetic fields. The measurements presented here were carried out at approximately 5 K and a steady photo current of a few µA was facilitated by a dc voltage from a battery source and white light from a halogen lamp. A lock-in amplifier provided the modulation signal at 1.23 kHz, and was also used to detect the EDMR signal from the device current.

The character of the EDMR spectra expected from Si:P are as for conventional ESR. An isolated donor electron, with $S = 1/2$ coupled to $I = 1/2$ $^{31}P$ nucleus, results in a hyperfine split doublet centred at $g = 1.9987$ (~3473 G in the spectra below) with ~42 G splitting. If donor pairs or clusters are present (exchange coupled), a central line may also exist. Charge traps also result in ESR lines. In particular an electron at an interface dangling bond ($P_b$ centre) is commonly observed (two overlapping broad lines at $g \sim 2.004$ and $2.008$ for the magnetic field $B_0$ parallel [110], often appearing as a single very broad line at $g = 2.0055$).

### 3.  Results

EDMR spectra collected for the three P concentrations and the various surface types are illustrated in Figs 2a, b and c. In all cases the maximum P and $P_b$ signals occur with the thermal oxide surface. This is perhaps unexpected given that the thermal oxide should have a much lower $P_b$ trap areal density than the native oxide. The results for the thermal oxide cases are reproduced for direct comparison in Fig 2d. Surprisingly there is little variation in the signals as a function of in P concentration, over nearly 3 orders of magnitude, and indeed are slightly smaller at the higher concentration ($10^{17}$ $cm^{-3}$).

Integrated signal strengths are estimated for comparison and summarised in table 1. The trend is clear with the P and $P_b$ signal magnitudes moving in concert. The devices with thermal oxide surfaces have the largest signals, while native oxides have the smallest (of the pre-prepared set). These trends are virtually independent of the P concentration. The H- and

D-terminated surfaces provide signals intermediate between the other two. These preparations, however, do not appear to represent true trap free interfaces when compared to the results in Fig 3, where much smaller signals were obtained from a freshly prepared (i.e. just prior to measurement) SiH surface.

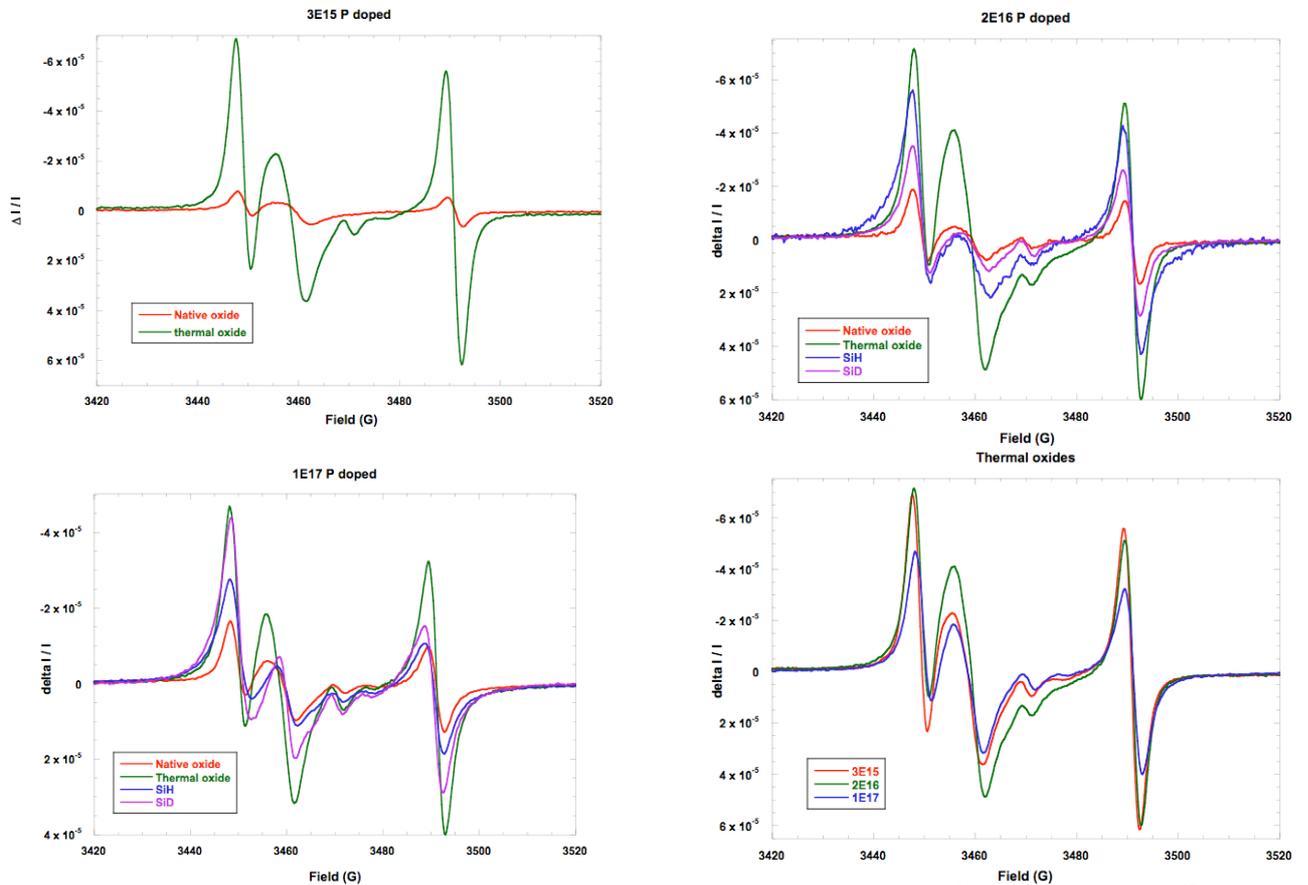

Fig. 2. (a), (b) and (c) are EDMR signals collected at ~5 K for Si:P devices with $3 \times 10^{15}$, $2 \times 10^{16}$ and $1 \times 10^{17}$ P $cm^{-3}$ respectively, for various surface types as labelled. The data in (d) are the thermal oxide results repeated for direct comparison.

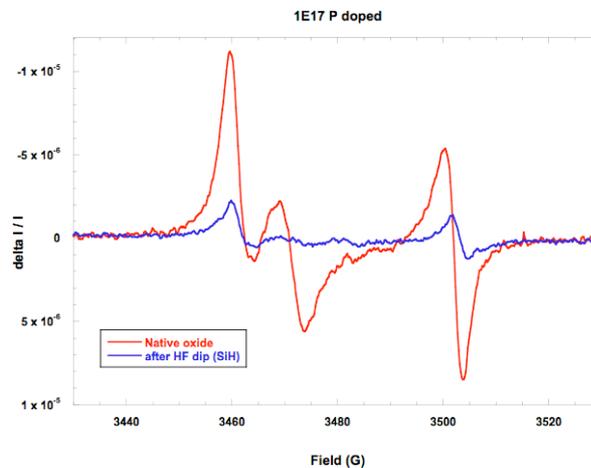

Fig. 3. EDMR signals collected at ~5 K for Si:P device with $1 \times 10^{17}$ P $cm^{-3}$, with native oxide surface and immediately following a HF etch which replaced the oxide with a hydrogen terminated surface.

Table 1. Parameters from least squares fits to the spectra.

| sample | sample parameters (concentrations in cm-3) | Signal (Arb. Units) | | |
|---|---|---|---|---|
| | | Pb | P doublet (total) | Centre |
| J1 | 1E17 doped native oxide | 1374 | 750 | 76 |
| N3 | 1E17 doped thermal oxide | 4366 | 3052 | 152 |
| J9 | 1E17 doped SiH | 1294 | 1390 | 87 |
| A12 | 1E17 doped SiD | 2345 | 2166 | 141 |
| D3 | 2E16 doped native oxide | 644 | 652 | 55 |
| V4 | 2E16 doped thermal oxide | 5905 | 3802 | 44 |
| D4 | 2E16 doped SiH | 1140 | 3536 | 44 |
| D11 | 2E16 doped SiD | 712 | 1842 | 57 |
| L4 | 3E15 doped native oxide | 427 | 378 | 0 |
| Y3 | 3E15 doped thermal oxide | 3084 | 3858 | 53 |
| J10 | 1E17 doped native oxide | 351 | 314 | 8 |
| J10 | 1E17 doped SiH (fresh) | 24 | 58 | 3 |

## 4. Discussion

These results clearly demonstrate that a photo current recombination model for EDMR in Si:P requiring the presence of deep charge traps is valid. There is an optimal relative concentration of traps to donors to get large EDMR resonances. Too many traps (eg native oxide) deplete donors. With too few traps (fresh SiH) the recombination path is blocked (NB our large area pre-prepared SiH and SiD were likely degraded between preparation and measurement c.f. the freshly dipped case). Anecdotally, from our measurements the thermal oxides in combination with bulk P densities between $10^{15}$ to $10^{16}$ cm$^{-3}$ seems optimal for maximum strength EDMR signals. Good thermal oxides on Si typically have an interface trap density of ~$10^{11}$ cm$^{-2}$ eV$^{-1}$ [5] which corresponds to an average spacing of about $3 \times 10^{-6}$ cm. While $3 \times 10^{15}$ and $2 \times 10^{16}$ P cm$^{-3}$ have average donor spacings of $7 \times 10^{-6}$ cm and $3 \times 10^{-6}$ cm respectively.

One other puzzle is the lack of variation in the P central line. Paired and clustered donors should be present in reasonable numbers only at higher ($1 \times 10^{17}$ cm$^{-3}$) P concentrations. It could be that some of this small signal is associated with regions of straggle near the implanted leads.

**Acknowledgments**
This work is supported by the Australian Research Council (ARC) and also in part by the National Security Agency (NSA) under Army Research Office (ARO) contract number W911NF-08-1-0527.

**References**
[1] G. Feher, 1959 *Phys. Rev.* **114** 1219-1244
[2] J. Schmidt and I. Solomon, 1966 *Compt. Rend. Paris* **263**, 169
[3] D.R. McCamey, H. Huebl, M.S. Brandt, W.D. Hutchison, J.C. McCallum, R.G. Clark and A.R. Hamilton, 2006 *Applied Physics Letters* **89** 182115-1 - 182115-3
[4] D. Kaplan, I. Solomon and N.F. Mott, 1976 *Journal de Physique – Letters* **41** 159
[5] S. Peterström, 1993 *Applied Physics Letters* **63**, 672